\documentclass{article}
\usepackage{amsmath}
\def\Ref#1{(\ref{#1})}
\usepackage{cite}
\begin{document}
\begin{titlepage}
\vspace*{10mm}
\begin{center}
{\large \bf Exclusion processes and boundary conditions}
\vskip 10mm
\centerline {\bf Farinaz Roshani $^{1,2,a}$ {\rm and} Mohammad Khorrami
$^{3,b}$}
\vskip 1cm
{\it $^1$ Institute for Advanced Studies in Basic Sciences,}
\\ {\it P. O. Box 159, Zanjan 45195, Iran}\\
{\it $^2$ Institute for Studies in Theoretical Physics and Mathematics,}
\\ {\it P. O. Box 5531, Tehran 19395, Iran}\\
{\it $^3$ Department of Physics, Alzahra University,}
\\ {\it Tehran, 19938-91167, Iran}\\
$^a$ farinaz@iasbs.ac.ir\\
$^b$ mamwad@mailaps.org\\

\end{center}
\vskip 2cm
\begin{abstract}
\noindent A family of boundary conditions corresponding to
exclusion processes is introduced. This family is a generalization
of the boundary conditions corresponding to the simple exclusion
process, the drop-push model, and the one-parameter solvable
family of pushing processes with certain rates on the continuum
\cite{AKK1,AKK2,RK}. The conditional probabilities are calculated
using the Bethe ansatz, and it is shown that at large times they
behave like the corresponding conditional probabilities of the
family of diffusion-pushing processes introduced in
\cite{AKK1,AKK2,RK}.
\end{abstract}
\vskip 2cm PACS numbers: 05.40.-a, 02.50.Ga
\end{titlepage}
\newpage
\section{Introduction}
In recent years, the asymmetric exclusion process and the problems
related to it, including for example the bipolimerization
\cite{4}, dynamical models of interface growth \cite{5}, traffic
models \cite{6}, the noisy Burgers equation \cite{7}, and the
study of shocks \cite{8,9}, have been extensively studied. The
dynamical properties of this model have been studied in
\cite{9,11}. As the results obtained by approaches like mean field
are not reliable in one dimension, it is useful to introduce
solvable models and analytic methods to extract exact physical
results. Among these methodes is the coordinate Bethe-ansatz,
which was used in \cite{Sc} to solve the asymmetric simple
exclusion process on a one-dimensional lattice. In \cite{AKK1}, a
similar technique was used to solve the drop-push model \cite{13},
and a generalized one-parameter model interpolating between the
asymmetric simple exclusion model and the drop-push model. In
\cite{AKK2}, this family was further generalized to a family of
processes with arbitrary left- and right- diffusion rates. All of
these models were lattice models. Finally, the behaviour of latter
model on continuum was investigated in \cite{RK}. The continuum
models of this kind are also investigated in \cite{SW1,SW2}. In
\cite{RK2}, another generalization of boundary conditions
corresponding to these processes was introduced, which shows an
exclusion process with annihilation as well as diffusion. In
\cite{RK3}, the Bethe-ansatz method was used to investigate
reaction-diffusion processes involving particles of more than one
type, on a lattice.

In the generalized model interpolating between the asymmetric
simple exclusion model and the drop-push model
\cite{AKK1,AKK2,RK}, there are two parameters $\lambda$ and $\mu$,
which control the pushing rate. Normalizing the diffusion rate to
one, it is seen that the sum of these two parameters should be one
to ensure the conservation of probability. These two parameters
appear in the boundary condition used instead of the reaction. The
question is that on continuum, what other kind of boundary
conditions, corresponding to other reactions, can be imposed. This
is what is investigated in the present paper.

The scheme of the paper is the following. In section 2, the
allowed boundary conditions are investigated. The criterion to
choose a boundary condition is the conservation of probability. It
is shown that the boundary condition is characterized by a
one-variable function which vanishes at the origin.

In section 3, the Bethe-ansatz solution for the $2$-particle
probability of this process is obtained, and its large-time
behavior is investigated. It is shown that the large-time behavior
of the system is determined by only the first nonzero derivative
of the function determining the boundary condition, at the origin.
If this function does have a nonvanishing linear term, then that
term determines the large-time behavior of the system, and the
results of \cite{RK} are recovered.

Finally, in section 4, boundary conditions are investigated for
which the first derivative of the function determining the
boundary condition vanishes at the origin. The mean position of
the particles, and the effective diffusion parameter are obtained.
\section{Boundary conditions}
Consider a collection of $N$ particles diffusing on a
one-dimensional continuum. So long as the particles do not
encounter each other, the master equation governing the
probability density of finding these at $x_1<\cdots < x_N$ is
\cite{RK}
\begin{equation}\label{1} {\partial\over{\partial
t}}P({\bf x};t)={1\over 2}\nabla^2 P({\bf x};t).
\end{equation}
In \cite{RK}, it was shown that the exclusion pushing condition
can be written as a boundary condition with two parameters (in
fact with only one independent parameter, since $\lambda +\mu
=1$):
\begin{equation}\label{2}
\left(\mu {\partial\over{\partial
x_i}}P-\lambda {\partial\over{\partial
x_{i+1}}}\right)P\Big\vert_{x_{i+1}=x_i}=0.
\end{equation}
To obtain this system, in \cite{RK} it was started with a
collection of particles diffusing on a one-dimensional lattice.
Each particle diffuses to the right with the rate one, if the
right site is free. If the right site is occupied, the particle
may push other particles and go to the right, with a special rate.
That is, we have the following process.
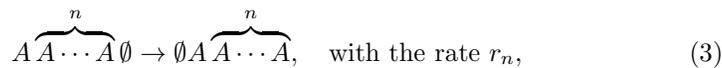
\begin{equation}\label{a1}
A\overbrace{A\cdots A}^n\emptyset\to\emptyset A\overbrace{A\cdots
A}^n,\quad\hbox{with the rate }r_n,
\end{equation}
where
\begin{equation}\label{a2}
r_n:=\left[1+\cdots+\left(\frac{\lambda}{\mu}\right)^n\right]^{-1}.
\end{equation}
It was shown in \cite{AKK1}, that such a system can be described
by the master equation
\begin{equation}\label{a3}
\dot P(\mathbf{x};t)=\sum_i
P(\mathbf{x}-\mathbf{e}_i;t)-N\,P(\mathbf{x};t),\quad
x_1<\cdots<x_N,
\end{equation}
subject to the boundary condition
\begin{align}\label{a4}
P(\dots,x_i=x,x_{i+1}=x,\dots;t)=&\lambda
P(\dots,x_i=x,x_{i+1}=x+1,\dots;t)\nonumber\\
+&\mu P(\dots,x_i=x-1,x_{i+1}=x,\dots;t).
\end{align}
In \Ref{a3}, $\mathbf{e}_i$ is a vector the components of which
are all zero, except for the $i$'th component, which is one. In
\cite{RK}, \Ref{1} and \Ref{2} were obtained as the limit of
\Ref{a3} and \Ref{a4} when the probability is a slowly-varying
function, with a suitable Galileo transformation, so that one has
diffusion to right and left (with equal rates). So, \Ref{1} and
\Ref{2} describe a system of particles diffusing on a continuum
and pushing each other with certain rates, as introduced in
\Ref{a1} and \Ref{a2}.

For the special case $N=2$, \Ref{1} and \Ref{2} become
\begin{equation}\label{3}
{\partial\over{\partial t}}P(x_1,x_2;t)={1\over
2}(\partial_1^2+\partial_2^2) P(x_1,x_2;t),
\end{equation}
and
\begin{equation}\label{4}
\mu {\partial_1}P(x,x)=\lambda {\partial_2}P(x,x).
\end{equation}
Using
\begin{equation}\label{5}
{\partial}P(x,x)={\partial_1}P(x,x)+{\partial_2}P(x,x),
\end{equation}
it is seen that the boundary condition \Ref{4} is equivalent to
\begin{equation}\label{6}
({\partial_1}-{\partial_2})P(x,x)=(\lambda-\mu){\partial}P(x,x).
\end{equation}

The passage from \Ref{a4} to \Ref{2} is not unique. But any
boundary condition has to ensure probability conservation.
Consider the evolution \Ref{3}, and note that it is valid only for
the physical region $x_1<x_2$. Integrating this over the physical
region, one arrives at
\begin{equation}\label{7}
{d\over{dt}}\int dx_1 dx_2\; P(x_1,x_2)={1\over 2}\int_{-\infty}
^{+\infty}dx (\partial_1-\partial_2)P(x,x).
\end{equation}
But using \Ref{6}, it is seen that the right-hand side of \Ref{7}
is zero. So,
\begin{equation}\label{8}
{d\over {dt}}\int dx_1 dx_2\; P(x_1,x_2)=0.
\end{equation}
This is the conservation of probability. However, the only thing
needed to ensure this conservation is that
$(\partial_1-\partial_2)P(x,x)$ be a total derivative of some
function with respect to $x$. So, one can in general use a
boundary condition like
\begin{align}\label{9}
(\partial_1-\partial_2)P(x,x)&=\alpha_1\partial
P(x,x)+\alpha_2\partial\partial P(x,x)+\cdots\nonumber\\
&=f(\partial)P(x,x),
\end{align}
where $f$ is an analytic function satisfying
\begin{equation}\label{10}
f(0)=0.
\end{equation}
Such a boundary condition ensures probability conservation. In
\cite{RK2}, a slightly different boundary condition was
introduced, which in this language is equivalent to $f(0)<0$. It
was seen there such a boundary condition describes annihilation as
well as diffusion.

\section{The Bethe-ansatz solution}
The Bethe-ansatz solution to the Master equation \Ref{1}, with the
boundary condition \Ref{9} is
\begin{equation}\label{b11}
P(\mathbf{x};t)=e^{Et}\Psi (\mathbf{x}),
\end{equation}
where $\Psi$ satisfies
\begin{equation}\label{b12}
E\Psi(\mathbf{x})=\frac{1}{2}\nabla^2\Psi(\mathbf{x}),
\end{equation}
and the boundary condition \Ref{9}. The solution to these, is
\begin{equation}\label{b13}
\Psi_{\mathbf{k}}(\mathbf{x})=\sum_{\sigma}A_\sigma\,e^{i\sigma(\mathbf{k})\cdot\mathbf{x}},
\end{equation}
with
\begin{equation}\label{b14}
E=-\frac{1}{2}\sum_i k_i^2,
\end{equation}
provided one can find $A_\sigma$'s so that \Ref{9} is satisfied.
The summation in \Ref{b13} is over the permutations of $N$
objects, Applying \Ref{9} to the case $x_j=x_{j+1}$, one obtains
\begin{equation}\label{b15}
i[k_{\sigma(j)}-k_{\sigma(j+1)}](A_\sigma-A_{\sigma\sigma_j})= i
[\tilde f(k_{\sigma(j)})+\tilde
f(k_{\sigma(j+1)})](A_\sigma+A_{\sigma\sigma_j}),
\end{equation}
where $\tilde f$ is defined as
\begin{equation}
\tilde f(k):=-i f(ik),
\end{equation}
and $\sigma_j$ is that permutation which changes $j$ to $j+1$ and
vice versa, and leaves other numbers unchanged. From \Ref{b15},
$A_\sigma$ can be obtained as
\begin{equation}\label{b16}
A_{\sigma\sigma_j}=S(k_{\sigma(j)},k_{\sigma(j+1)}) A_\sigma,
\end{equation}
where
\begin{align}\label{b17}
S(k_1,k_2)&={{k_2-k_1+\tilde f(k_1+k_2)}\over{k_2-k_1-\tilde
f(k_1+k_2)}}\nonumber\\
&=1+{{2\tilde f(k_1+k_2)}\over{k_2-k_1-\tilde f(k_1+k_2)}}.
\end{align}
As $\sigma_1\sigma_2\sigma_1=\sigma_2\sigma_1\sigma_2$, the
coefficients $A$, should satisfy the consistency condition
\begin{equation}\label{bb18}
A_{\sigma_1\sigma_2\sigma_1}=A_{\sigma_2\sigma_1\sigma_2}.
\end{equation}
The above criterion, in terms of $S$ is
\begin{equation}\label{bb19}
S(k_2,k_3)S(k_1,k_3)S(k_1,k_2)=S(k_1,k_2)S(k_1,k_3)S(k_2,k_3),
\end{equation}
which is obviously an identity.

Using this Bethe-ansatz solution, the conditional probability can
be written as
\begin{equation}\label{b18}
P(\mathbf{x};t|\mathbf{y};0)=\int\frac{dk^N}{(2\pi)^N}\Psi_{\mathbf{k}}(\mathbf{x})
\,e^{E(\mathbf{k})t-i\mathbf{k}\cdot\mathbf{y}},
\end{equation}
where in $\Psi_{\mathbf{k}}$, the coefficient of
$e^{i\mathbf{k}\cdot\mathbf{x}}$ is set to be equal to one.

This shows the integrability of the system, in the sense that the
$N$-particle scattering matrix ($A_\sigma$'s) can be expressed as
a product of two-particle scattering matrices.

\subsection{The conditional probability for the 2-particle sector}
The conditional probability for the 2-particle sector, is written
as
\begin{equation}\label{16}
P({\bf x};t|{\bf y};0)={1\over 4\pi^2}\int dk_1 dk_2 \left[
e^{ik_1x_1+ik_2x_2}+S(k_1,k_2)e^{ik_1x_2+ik_2x_1} \right]
e^{Et-i{\bf k}\cdot{\bf y}},
\end{equation}
so that it satisfies \Ref{3} and the the initial condition
\begin{equation}\label{17}
P({\bf x},0|{\bf y},0)=\delta(x_1-y_1)\delta(x_2-y_2)
\end{equation}
for the physical region ($x_1<x_2,\quad y_1<y_2$).

Using the form of $S$, one can write this conditional probability
as
\begin{align}\label{18}
P({\bf x};t|{\bf y};0)=&\int{{d^2k}\over{4\pi^2}} e^{Et-i{\bf
k}\cdot{\bf y}}\Bigg\{ e^{i(k_1x_1+k_2x_2)}\nonumber\\
&+\left[1+{{2\tilde f(k_1+k_2)}\over{k_2-k_1-\tilde
f(k_1+k_2)}}\right] e^{i(k_1x_2+k_2x_1)}\Bigg\}.
\end{align}
In the above integration, the possible ambiguity arising from the
the pole of the fraction is removed through $k_1\to
k_1+i\,\varepsilon$ and $k_2\to k_2-i\,\varepsilon$, where the
$\varepsilon\to 0^+$ limit of the integral is meant. This ensures
that the probability tends to zero as $x_1\to -\infty$ or $x_2\to
+\infty$.

The two first integrals of the right-hand side are easily
calculated. So,
\begin{equation}\label{19}
P({\bf x};t|{\bf y};0)={1\over{2\pi t}}\left\{
e^{-[(x_1-y_1)^2+(x_2-y_2)^2]
/(2t)}+e^{-(z_1^2+z_2^2)/(2t)}\right\} +I_3,
\end{equation}
where
\begin{equation}\label{20}
z_1:=x_1-y_2,\qquad z_2:=x_2-y_1.
\end{equation}
To obtain the third integral ($I_3$), one uses the change of
variable
\begin{equation}\label{21}
k:=k_1+k_2,\qquad q:=k_2-k_1-\tilde f(k_1+k_2).
\end{equation}
Then,
\begin{align}\label{22}
I_3=&\int{{dkdq}\over{4\pi^2}}{{\tilde f(k)}\over q} e^{-t\{
k^2+[q+\tilde f(k)]^2\} /4}e^{i\{k A_2+[q+\tilde f(k)]A_1\}
/2}\nonumber\\ =&\sqrt{\pi\over
t}\int{{dk}\over{4\pi^2}}f(ik)e^{[ikA_2+A_1f(ik)]/2}
e^{-tk^2/4}\nonumber\\ &\times \int_{-\infty}^{A_1}dA\;
e^{-[A^2+2tA f(ik)]/(4t)},
\end{align}
where
\begin{equation}\label{23}
A_1:=z_1-z_2,\qquad A_2:=z_1+z_2.
\end{equation}
Simplifying \Ref{22}, one arrives at
\begin{align}\label{24}
I_3=&\sqrt{\pi\over
t}\int{{dk}\over{4\pi^2}}\int_{-\infty}^{A_1}dA\;
f\left(2{\partial\over{\partial A_2}}\right)e^{\displaystyle{
{{A_1-A}\over 2}f\left(2{\partial\over{\partial
A_2}}\right)}}\nonumber\\ &\times
e^{-\displaystyle{{tk^2-2ikA_2}\over 4}}
e^{-\displaystyle{{A^2}\over{4t}}}\nonumber\\ =&{1\over{2\pi
t}}\int_{-\infty}^{A_1}dA\; f\left(2{\partial\over{\partial
A_2}}\right)e^{\displaystyle{ {{A_1-A}\over
2}f\left(2{\partial\over{\partial A_2}}\right)}}
e^{-\displaystyle{{A^2+A_2^2}\over{4t}}}.
\end{align}
To show that \Ref{19} does indeed satisfy the initial conditions,
first note that
\begin{equation}\label{a11}
P({\bf x};0|{\bf y};0)=\delta(x_1-y_1)\,\delta(x_2-y_2) +
\delta(x_1-y_2)\,\delta(x_2-y_1) +I_3(t=0).
\end{equation}
The second term is obviously zero in the physical region $x_1<x_2$
and $y_1<y_2$. For the third term, we have from the first equality
in \Ref{22}
\begin{equation}\label{a12}
I_3(t=0)=\int\frac{dk}{4\pi^2}\,\tilde f(k) e^{i[k A_2+A_1\tilde
f(k)]/2} \int\frac{dq}{q-i\varepsilon}\,e^{iqA_1/2}.
\end{equation}
The integral over $q$ is, however, zero in the physical region.
Since in the physical region $A_1<0$, and one can close the
integration contour by adding a large semicircle in the lower half
plane of $q$, to the real line. There are no singularities inside
the closed contour, so the integral vanishes. So \Ref{19} does
indeed satisfy the initial conditions.

Now consider the large-time behavior of the system. using the
change of variable
\begin{equation}\label{25}
A_i=:2a_i\sqrt{t},
\end{equation}
the integral $I_3$ becomes
\begin{equation}\label{26}
I_3={1\over{\pi t}}\int_{-\infty}^{a_1}da\;\sqrt{t}
f\left({1\over\sqrt{t}}{\partial\over{\partial
a_2}}\right)e^{\displaystyle{
\sqrt{t}(a_1-a)f\left({1\over\sqrt{t}}{\partial\over{\partial
a_2}}\right)}} e^{-(a^2+a_2^2)}.
\end{equation}
It is seen that at large times, the dominant term comes from the
first nonzero term in the expansion of $f$. This
means that if the first derivative of $f$ does not vanish at the
origin, then at large times, $f$ is equivalent to a linear
function. But a linear $f$ is just the boundary condition used in
\cite{RK}. So at large times, the boundary condition has
effectively only one free parameter determining the interaction.

If $f$ is an at-most-quadratic polynomial, then the integration
over $k$ in \Ref{22} can be done. For the simple example
\begin{equation}\label{28}
f(x)=\alpha_2 x^2,
\end{equation}
the result would be
\begin{align}\label{29}
I_3=&\alpha_2\int_{-\infty}^{A_1}{{dA}\over{2\pi}}\left\{
{{A_2^2}\over{[t+2\alpha (A-A_1)]^2}}-{2\over{t+2\alpha
(A-A_1)}}\right\}\nonumber\\
&\times{1\over{\sqrt{t[t+2\alpha
(A-A_1)]}}}e^{-{{A^2}\over{4t}}-{{A_2^2}\over{4t+8\alpha
(A-A_1)}}}.
\end{align}
\section{Boundary conditions corresponding to functions $f$ with
vanishing first derivative at the origin} Taking the form
\begin{equation}\label{30}
f(x)=\alpha_n x^n+O(x^{n+1})
\end{equation}
for $f$ (where $n>1$), it is seen from \Ref{26} that at large
times, the leading term of $I_3$ is
\begin{equation}\label{31}
I_3\sim{\alpha_n\over{\pi\sqrt{t^{n+1}}}}\int_{-\infty}^{a_1}da
{{\partial^n}\over{\partial a_2^n}} e^{-(a^2+a_2^2)}.
\end{equation}
The integration in the left-hand side, can now be easily done, and
one arrives at
\begin{equation}\label{32}
I_3\sim{\alpha_n\over{2\sqrt{\pi t^{n+1}}}}[1+{\rm erf}(a_1)]
{{\partial^n}\over{\partial a_2^n}} e^{-a_2^2}.
\end{equation}

Two important quantities to be calculated, are the mean position
of the particles, and the variance of the mean position. The
expectation value of any function $g(x_1,x_2)$, is obtained
through
\begin{align}\label{33}
\langle g(x_1,x_2)\rangle:=&\int_{x_1\leq x_2} d^2x\; P({\bf
x}|{\bf y})g(x_1,x_2),\nonumber\\
=&\langle g(x_1,x_2)\rangle_0+\langle g(x_1,x_2)\rangle_3,
\end{align}
where
\begin{align}\label{34}
\langle g(x_1,x_2)\rangle_3:=&\int_{x_1\leq x_2} d^2x\;
I_3(x_1,x_2)g(x_1,x_2),\nonumber\\
=&2t\int_{-\infty}^{\infty}da_2\int_{-\infty}^{(y_1-y_2)/(2\sqrt{t})}
da_1\;I_3(x_1,x_2)g(x_1,x_2),
\end{align}
and
\begin{align}\label{35}
\langle g(x_1,x_2)\rangle_0:=&\int_{x_1\leq x_2} d^2x\;
{1\over{2\pi t}}
e^{-[(x_1-y_1)^2+(x_2-y_2)^2]/(2t)}g(x_1,x_2),\nonumber\\
&+\int_{x_1\geq x_2} d^2x\; {1\over{2\pi t}}
e^{-[(x_1-y_1)^2+(x_2-y_2)^2]/(2t)}g(x_2,x_1).
\end{align}
If $g(x_1,x_2)=g(x_2,x_1)$, then the above expression becomes
simpler:
\begin{equation}\label{36}
\langle g(x_1,x_2)\rangle_0=\int d^2x\; {1\over{2\pi t}}
e^{-[(x_1-y_1)^2+(x_2-y_2)^2]/(2t)}g(x_1,x_2).
\end{equation}

Using these, we want to calculate
\begin{equation}\label{37}
X:={1\over 2}\langle x_1+x_2\rangle,
\end{equation}
and
\begin{equation}\label{38}
(\Delta X)^2:={1\over 2}\langle x_1^2+x_2^2\rangle -X^2.
\end{equation}
Using \Ref{36}, it is easily seen that
\begin{equation}\label{39}
X_0={{y_1+y_2}\over 2},
\end{equation}
and
\begin{equation}\label{40}
{1\over 2}\langle x_1^2+x_2^2\rangle_0=(X_0)^2+t.
\end{equation}

From \Ref{26}, it is seen that $I_3$ is the $n$'th derivative of
some function with respect to $a_2$. So, if $g$ is a polynomial of
order less than $n$ with respect to $a_2$, the right-hand side of
\Ref{34} vanishes; this can be seen through $n$ times integration
by parts. The functions relevant to the calculation of the mean
position and the variance, in terms of $a_1$ and $a_2$, are
polynomials of $a_1$ and $a_2$ of at most the second degree. From
\Ref{30}, it is seen then that only $\langle a_2^2\rangle_3$ may
be nonvanishing, and that even this term vanishes if $n>2$. So,
for $n>2$ we have $X=X_0$ and ${\rm Var}={\rm Var}_0$. For $n=2$,
it is seen that $I_3$ contains second and higher derivatives of
functions with respect to $a_2$. In calculating $\langle
a_2^2\rangle_3$, only the second derivatives are relevant. That
is, one can write
\begin{equation}\label{41}
I_3={\alpha_2\over{2\sqrt{\pi t^3}}}[1+{\rm erf}(a_1)]
{{\partial^2}\over{\partial a_2^2}} e^{-a_2^2}+\widetilde{I_3},
\end{equation}
where $\widetilde{I_3}$ is irrelevant to the calculation of
$\langle a_2^2\rangle_3$. So, one arrives at
\begin{equation}\label{42}
\langle a_2^2\rangle_3={{2\alpha_2}\over{\sqrt{t}}}
\int_{-\infty}^{(y_1-y_2)/(2\sqrt{t})}da_1\;[1+{\rm erf}(a_1)].
\end{equation}
Using
\begin{equation}\label{43}
x_1^2+x_2^2={1\over 2}(A_1^2+A_2^2)+B,
\end{equation}
where $B$ is a first order polynomial, one arrives at
\begin{align}\label{44}
{1\over 2}\langle x_1^2+x_2^2\rangle_3&=t\langle a_2^2\rangle_3
\nonumber\\
&=2\alpha_2\sqrt{t}\int_{-\infty}^{(y_1-y_2)/(2\sqrt{t})}da_1.
\end{align}
From this, \Ref{40}, and the fact the mean does not depend on $f$
(so long as the first derivative of $f$ vanishes at the origin) it
is seen that
\begin{equation}\label{45}
{\rm
Var}=t+2\alpha_2\sqrt{t}\int_{-\infty}^{(y_1-y_2)/(2\sqrt{t})}
da_1\;[1+{\rm erf}(a_1)].
\end{equation}
This becomes even simpler at large times. At large times, the
upper limit in the above integral tends to zero. So, using
\begin{equation}\label{46}
\int_{-\infty}^0[1+{\rm erf}(x)]dx={1\over{\sqrt{\pi}}},
\end{equation}
one arrives at
\begin{equation}\label{47}
{\rm Var}=t+2\alpha_2\sqrt{t/\pi}+O(t^{-1/2}).
\end{equation}
The effective diffusion parameter is defined as the derivative of
the above:
\begin{equation}\label{48}
{{d{\rm Var}}\over{dt}}=1+{\alpha_2\over{\sqrt{\pi
t}}}+O(t^{-3/2}).
\end{equation}
It is seen that as $t\to\infty$, this parameter tends to one, in
agreement with \cite{Sc,RK}.

\end{document}